 \documentclass[final,5p,times,twocolumn]{elsarticle}

\usepackage{amssymb}
\usepackage[svgnames]{xcolor}
\usepackage{amsmath}
\usepackage{multirow}
\usepackage{placeins}
\usepackage{bigints}
\usepackage[bookmarks=false]{hyperref}
\hypersetup{colorlinks,linkcolor={blue},citecolor={blue},urlcolor={red}}
\graphicspath{{Figures/}}

\begin{document}

\begin{frontmatter}



\title{Utilizing polydispersity in composite fibrous based sound absorbing materials}


\author[inst1,inst2,inst3,inst4]{Quang Vu Tran}

\affiliation[inst1]{organization={Univ Gustave Eiffel, Univ Paris Est Creteil},
            addressline={CNRS, UMR 8208, MSME}, 
            postcode={F-77454}, 
            state={Marne-la-Vallée},
            country={France}}

\author[inst1,inst4]{Camille Perrot}
\author[inst2,inst4]{Raymond Panneton}
\author[inst3]{Minh Tan Hoang}
\author[inst3]{Ludovic Dejaeger}
\author[inst3]{Valérie Marcel}
\author[inst3]{Mathieu Jouve}

\affiliation[inst2]{organization={Département de Génie Mécanique},
            addressline={Université de Sherbrooke}, 
            postcode={J1K 2R1}, 
            state={Québec},
            country={Canada}}
\affiliation[inst3]{organization={Adler Pelzer Group, Acoustic TechCenter R$\&$D},
            addressline={ Z.I. François Sommer – BP13}, 
            postcode={08210}, 
            state={Mouzon},
            country={France}}
\affiliation[inst4]{email={quang.vu.tran@usherbrooke.ca, camille.perrot@univ-eiffel.fr, raymond.panneton@usherbrooke.ca}}

\begin{abstract}
The distribution of fiber diameters plays a crucial role in the transport and sound absorbing properties of a three-dimensional random fibrous (3D-RF) composites. Conventionally, volume-weighted averaging of fiber diameters has been utilized as an appropriate microstructural descriptor to predict the static viscous permeability of 3D-RF composites. However, the long wavelength acoustical properties of a 3D-RF composites are also sensitive to the smallest fibers, this is particularly true in the high-frequency regime. In our recent research, we demonstrated that an inverse volume-weighted averaging of fiber diameters can effectively serve as a complementary microstructural descriptor to capture the high-frequency behavior of polydisperse fibrous media. In the present work, we review the identification of two representative volume elements (RVEs) which relies on the reconstruction of 3D-RF composites having volume-weighted and inverse-volume weighted averaged fiber diameters, respectively in the low-frequency and high frequency regimes. We examine the implication of such a weighting procedure on the transport and sound absorbing properties of polydisperse fibrous media, highlighting their potential advantages. Furthermore, we discuss the challenges associated with this research field. Finally, we provide a brief perspective of the future directions and opportunities for advancing this area of study, aiming to overcome challenges and extend the benefits of employing polydispersity as a new lever for the optimization of 3D-RF composites in sound-absorbing materials.
\end{abstract}




\end{frontmatter}


\section{Introduction}
\label{sec:Introduction_a2}
Three-dimensional random fibrous (3D-RF) composites have become one of the most widespread technologies in large-scale manufacturing of sound-absorbing materials known as nonwovens or felts, employing various types of fibers to achieve their acoustical constraints. Some commonly used manufacturing processes include the airlay process \cite{BHAT2007143, gramsch2016aerodynamic} and the carding, cross-lapping, and needing process \cite{AHMED2007368, albrecht2006nonwoven}. Traditional thermobonded airlaid nonwoven typically incorporates a mixture of non-homogeneous shoddy fibers ($75\%$) and bicomponent fibers (25\%) in the airlay process.

The bicomponent fibers have a core made from PET and a surface made from coPET. On the other hand, to facilitate recycling, needlefelt nonwoven use a mixture of PET fibers ($60\%$) and bicomponent fibers ($40\%$). In both cases, in post-processing, the nonwoven materials are reinforced by thermobonding (with a chosen compression ratio), where the bicomponent fibers have an adhesive effect. However, the use of non-homogeneous fibers in the manufacturing process comes with some drawbacks. One of the disadvantages is that the transport mechanisms can be significantly impacted by the fiber size variations induced by the manufacturing process. For example, shoddy fibers are obtained after the tearing of textile waste from a mixture of $55\%$ cotton and $45\%$ PET, and cotton fibers can introduce fiber dust particles, which affect the distribution of fiber diameters. The manufacturing process of nonwoven leads to a wide distribution of fiber diameters and fiber orientations, which affects the frequency-dependent response of the fluid within the corresponding 3D-RF composites. Moreover, mixing fibers of different sizes is generally recommended. The largest fibers ensure the rigidity and mechanical stability of the nonwoven material, while the smallest fibers provide sound absorbing capabilities to the fibrous materials. Therefore, an important research direction in this field is the study of various distributions of fiber diameters and orientations and the development of models that account for the corresponding frequency-dependent response functions. Several approaches have been explored to achieve this goal. Perhaps the most direct method is to conduct a series of laboratory measurements on samples of varying fiber size and orientation \cite{lei2018prediction,TANG2017360}. Alternatively, in the quest for theoretical understanding, one may seek to better understand the mathematical \cite{  auriault1980dynamic, burridge1981poroelasticity, tarnow1996airflow, umnova2000cell,semeniuk2017microstructure, VENEGAS2021109006} or physical \cite{johnson1987theory, lafarge1997dynamic} basis of the generalized Darcy-scaled equations for macroscopic transport. Lastly, one can consider studies based on numerical simulation \cite{martys1992length, koponen1998permeability,OPIELA2020107833}. Notably, Peyrega and Jeulin \cite{Peyrega2013} used a volume-weighted average diameter to successfully predict the static viscous permeability of heterogeneous fibrous materials and our group has recently demonstrated that an inverse volume weighted average diameter can serve as a complementary microstructural descriptor to predict the characteristic lengths and tortuosity of these polydisperse 3D random fibrous materials.

In this paper, we aim to provide a comprehensive overview of research advancements in employing 3D polydisperse RF microstructures, with a particular emphasis on using polydispersity as a new lever of optimization. Starting from existing specimens of nonwoven fibrous materials, we focus specifically on three different types of studies: (i) a parametric study, using the coefficient of variation of the fiber diameters (polydispersity degree, $P_d$) as input parameter; (ii) optimizing the sound absorption average at normal incidence over a wide frequency range of a given nonwoven specimen by controlling the degree of polydispersity of the fiber diameters; (iii) achieving targets of industrial interest consisting of lowering the lower frequency from which the sound absorption in a diffuse field is greater than $80\%$. 

The paper is organized in the following sequences. In the first section, we delve into the fundamental transport properties of 3D-RF composites used as sound-absorbing materials that are characterized by a wide distribution of their fiber diameters. We discuss the unique transport properties and microstructural descriptors derived from the polydispersity of these materials, highlighting their potential advantages over traditional fibrous materials. This section serves as a foundation for understanding the subsequent discussions. Building upon the understanding of 3D microstructure based RF materials, we dedicate the second section to exploring the advantages offered by employing specific polydispersity degree as a microstructural optimization lever. We discuss the benefits associated with tuning the polydispersity degree of 3D-RF composites in terms of their transport properties, sound absorption average at normal incidence, sound absorption targets in diffuse field. By doing so, we highlight our contribution to the field, showing the advancements we have made in leveraging these materials for polydisperse fiber diameters. Finally, we conclude the article with a brief perspective and outlook on current challenges and the future of research in this area. We discuss emerging challenges, potential directions for further explorations, and the significance of continued investigations into 3D-RF composite materials as a microstructural optimization lever of their physical properties. By summarizing the current state of the field and presenting potential avenues for future research, we aim to inspire and guide researchers in this exciting area of study.
\FloatBarrier
\section{Nonwoven fibrous media characterized by local heterogeneity}
\subsection{Predicting transport properties}
Martys and Garboczi \cite{martys1992length} demonstrated through numerical computations the important effect spatial randomness in the pore space has on flow problems. In particular, they established a crucial distinction between the electric fields and the fluid-flow fields for a given pore structure. Leveraging this distinction, they identified that in a random pore structure with a distribution of pore sizes, the fluid-flow will tend to go more through the largest pore necks; while the electrical current-flow rate was comparatively less sensitive to the width of narrow necks. As a result, the fluid velocity fields and the electric fields can sample the pore space quite differently. Subsequently, Peyrega and Jeulin \cite{Peyrega2013} made substantial advances in this area, particularly in identifying the volume-weighted average radius as a key microstructural descriptor for the determination of the representative volume element of a random fibrous medium. Moreover,  Tran \cite{tran2023effect} and collaborators delved into the associated microscopic basis of macroscopic viscous permeability, shedding further light on how macroscopic viscous permeability depends on the largest pore sizes of a locally heterogeneous 3D-RF composite. The volume-weighed average diameter $D_v$ of a wide distribution of fiber diameters is larger than non-weighted average diameter $D_m$, and the number of introduced fibers in a cubic box of size $L/D_m$ is therefore lower in the volume-weighted average diameter case than in the non-weighted average diameter one -- while the porosity is known and given by measurements in both cases. As a result, the reconstructed pore sizes of 3D-RF composites having a volume-weighted average diameter match with the largest pore necks through which the viscous fluid-flow will tend to go in the corresponding nonwoven specimen. This reconstruction procedure enables the identification of a REV having $D_v$ as the average diameter and the computed viscous permeability $k_0$ to be consistent with the measured values [$S_D=45 mm$, $L/D_m\approx25 (-)$, $L=400 \mu m$] in real samples whose sample diameters are much larger than those of the RVE \cite{tran2023effect}.
\begin{figure*}[ht]
    \centering
    \includegraphics[width=12cm]{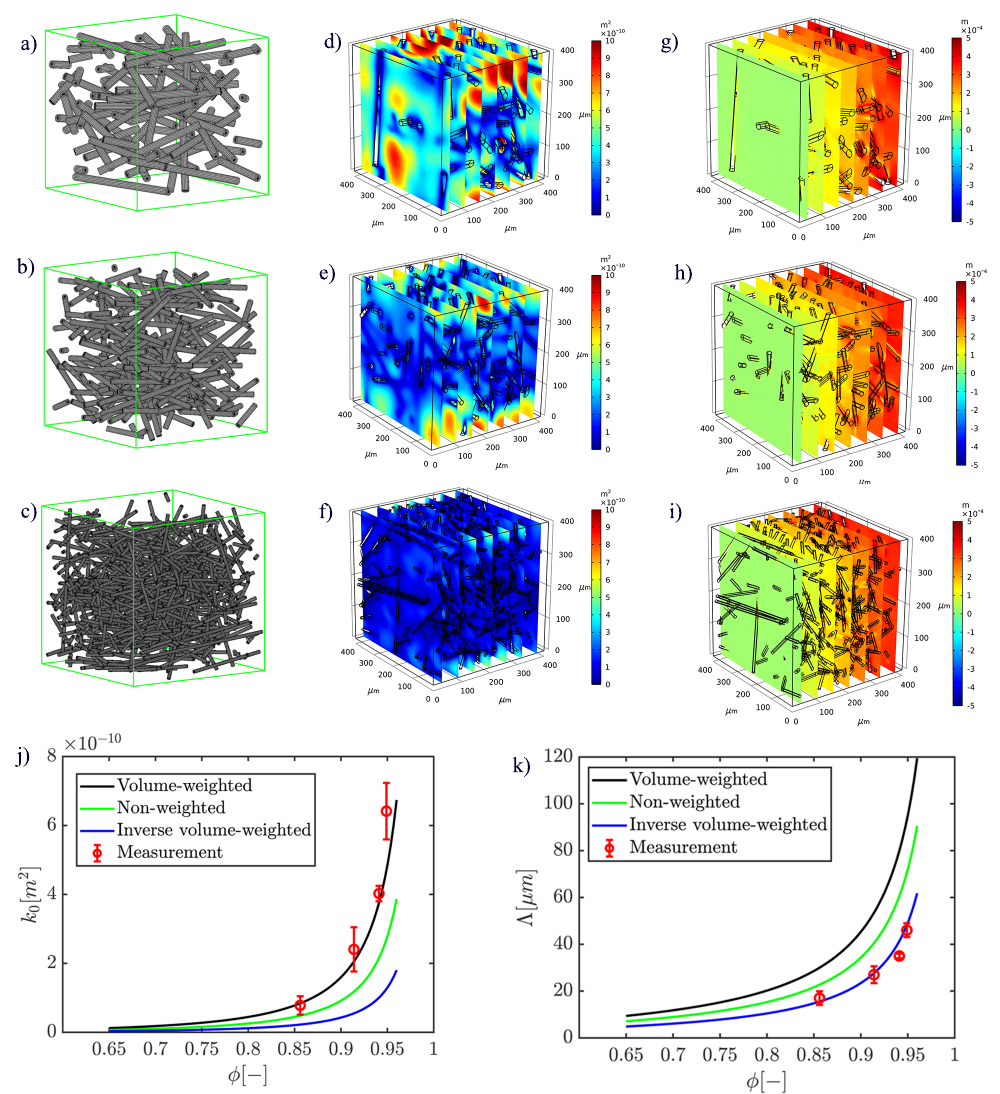 }
    \caption{Effect of polydispersity on the transport properties. (a)-(c) Three-dimensional random fibrous microstructures corresponding to the sample denoted F2 using volume-weighted, non-weighted, inverse volume-weighted average diameters, respectively. (d)-(f) Velocity fields expressed as local permeability ($k_{0zz}[m^2]$) corresponding to Stokes flow in the $z$ direction with the REV reconstructed using (a) volume weighted, (b) non-weighted and (c) inverse volume-weighted average diameter. (g)-(i) Scaled potential field $(\varphi [m]$ corresponding to potential flow in the $z$ direction [-] with the REV reconstructed using  (a) volume weighted, (b) non-weighted and (c) inverse volume-weighted average diameter. (j) The static viscous permeability $k_0 (m^2)$ of nonwoven fibrous materials (F1, F2, F3, F4) as a function of porosity $\phi$(-). (k) The viscous characteristic length $\Lambda(\mu m)$ of nonwoven fibrous materials (F1, F2, F3, F4) as a function of porosity $\phi$(-).}
    \label{fig:figure1_article2}
\end{figure*}

The combined presence of a wide distribution of fiber diameters and random microstructure promotes pore-size local heterogeneity, leading to the identification of the volume-weighted average diameter, as being an appropriate microstructural descriptor for the prediction of the static viscous permeability. The use of the finite element method revealed that velocity fields are consistent with experimental measurements of viscous permeability using volume-weighted average diameter $D_v$, as shown in Fig. \ref{fig:figure1_article2}(a)-\ref{fig:figure1_article2}(d) and \ref{fig:figure1_article2}(j). However, the electric current paths are clearly less concentrated and tortuous than do the fluid-flow paths, so there are significantly fewer stagnant areas for the electrical current flow than for the fluid flow. $\Lambda$ is defined \cite{johnson1986new} by the following ratio of integrals: $\Lambda=2\int{\mid\textbf{E}(\textbf{r})\mid^2dV_p}/\int{\mid\textbf{E}(\textbf{r})\mid^2dS}$, where $\textbf{E}(\textbf{r})$ is the magnitude of the electrical field in the pore space, $dV_p$ is the volume element in the pore space, and $dS$ is the surface element on the pore solid interface. Therefore, $\Lambda$ can be thought of as a dynamically weighted hydraulic radius \cite{johnson1987theory}, where the weighting procedure substantially favors the smaller pores because of current conservation. Fig. \ref{fig:figure1_article2}(c) shows a 3D-RF composite where inverse volume-weighted average diameter is used as the appropriate microstructural descriptor for the prediction of the viscous characteristic length $\Lambda$. The geometrical reconstruction method is given in our previous work \cite{tran2023effect}. The smaller pores of the nonwoven polydisperse fibrous materials were probed by measuring the experimental value of the $\Lambda$ parameter, a value captured by using inverse volume-weighted average diameter $D_{iv}$ during the reconstruction procedure, as shown in Figs. \ref{fig:figure1_article2}(i) and \ref{fig:figure1_article2}(k). The change in pore size reconstruction induced by fiber diameter polydispersity and volume-based fiber diameter weighting is shown in Fig. \ref{fig:figure1_article2}(a)-\ref{fig:figure1_article2}(c). At the scale of the reconstructed unit-cell, the non-weighted average diameter $D_m$ was neither an appropriate microstructural descriptor for describing the largest pores connected with viscous fluid flow nor a correct one to identify the smallest dynamically connected pore sizes probed by electrical current flow; Fig. \ref{fig:figure1_article2}(b), \ref{fig:figure1_article2}(e) and \ref{fig:figure1_article2}(h). $D_v$ based and $D_{iv}$ based REVs predictions all showed excellent agreement respectively with static viscous permeability $k_0$ and characteristic length $\Lambda$ measurements over a large range of porosities $\phi$; Fig. \ref{fig:figure1_article2}(j) and \ref{fig:figure1_article2}(k).
\FloatBarrier
\subsection{Polydispersity effect on the transport properties of three-dimensional random fibrous microstructures}
\begin{figure*}[ht]
    \centering
    \includegraphics[width=12cm]{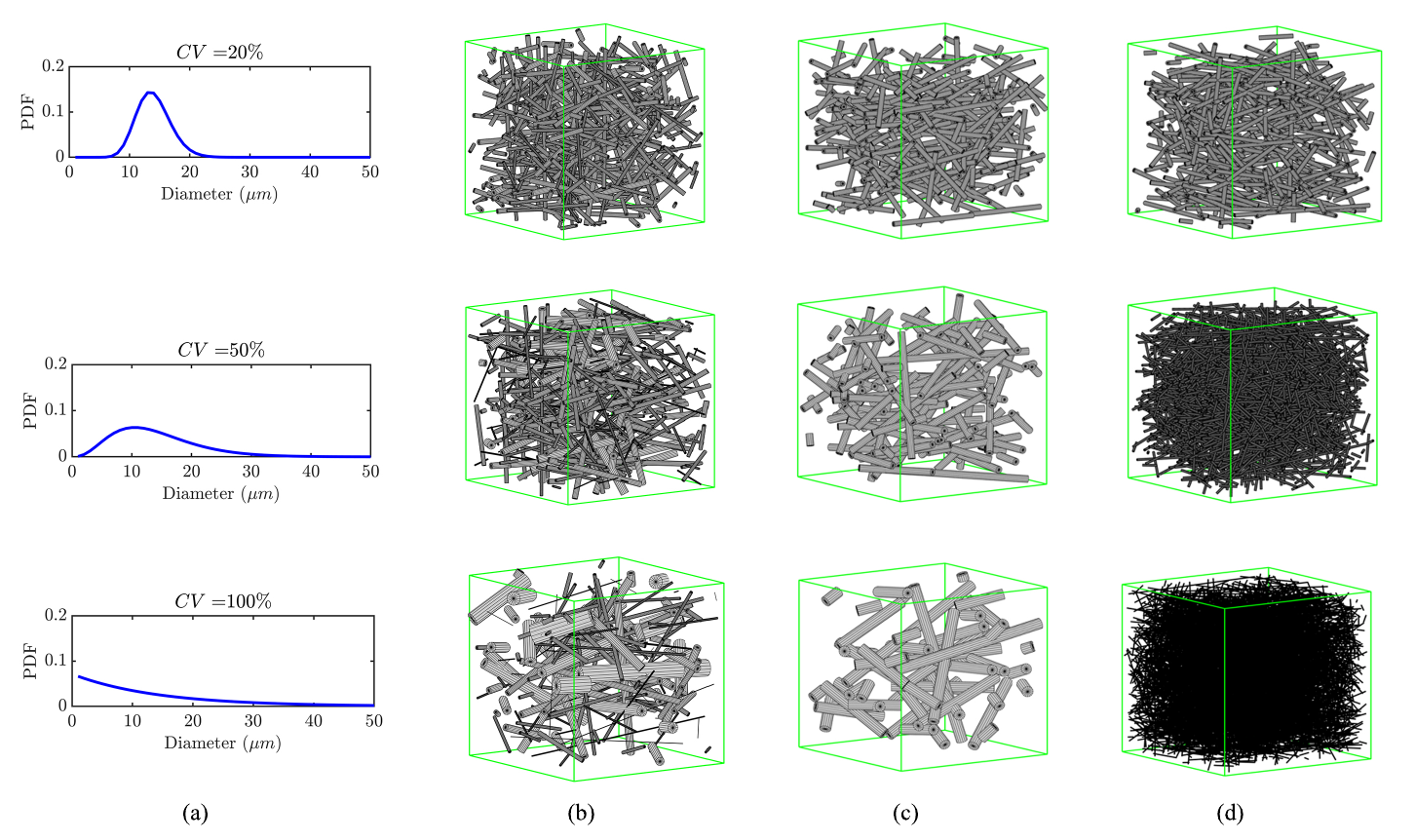}
    \caption{Polydispersity effect on the representative elementary volumes of three-dimensional random fibrous microstructures. (a) The probability distribution function of Gamma-based fiber diameters ($D_m=14 \mu m$) for increasing coefficients of variation $CV$ of fiber diameters ($CV=20\%$, $CV=50\%$, $CV=100\%$). (b) Geometric reconstruction of corresponding three-dimensional random fibrous microstructures ($\phi=94.1\%$). Geometric reconstructions of (c) $D_v$-based and (d) $D_{iv}$-based representative elementary volumes used to predict the viscous fluid flow and electrical current flow effective transport properties of polydisperse fibrous media.}
    \label{fig:figure2_article2}
\end{figure*}

\begin{figure}[ht]
    \centering
    \includegraphics[width=8.5cm]{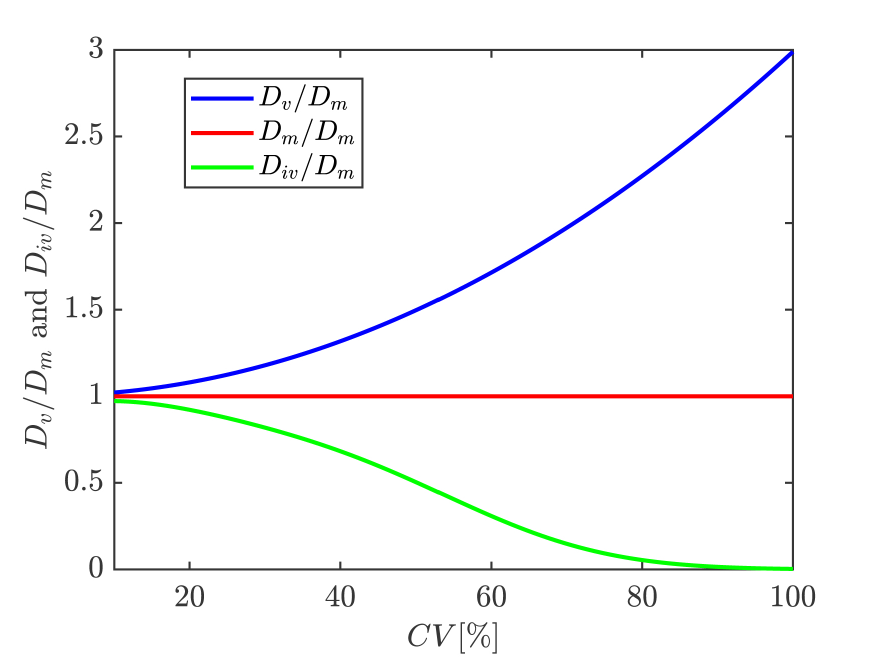}
    \caption{Evolution of the volume-weighted average diameter $D_v$ and the inverse volume-weighted average diameter $D_{iv}$ as a function of the polydispersity degree $P_d$ of a fiber diameters' distribution quantified through the coefficient of variation $CV$ of a Gamma-based law at constant mean diameter $D_m$.}
    \label{fig:Dviv_CV}
\end{figure}

Polydispersity effect occurs when particles of varied sizes in the dispersed phase modifies the expected behavior of the corresponding system. The wide distribution of fiber diameters changes the transport parameters of a 3D-RF composite characterized by a sharply peaked distribution of fiber diameters, resulting is a non-conventional REV. Our previous works describe the Gamma-based distribution of fiber diameter configuration and confirms this effect in glass wool, cotton felts and PET felts' composites.\cite{he2018multiscale, tran2023effect} Figure \ref{fig:figure2_article2}(a) illustrates the Gamma distribution shape of fiber diameters according to the equation
\begin{align}
    f(D; a, b) = \frac{1}{\Gamma({a})b^{a}}D^{a-1}e^{-\frac{D}{b}}, \label{eq:gamma pdf}
\end{align}
where $D$ is the fiber diameter of the probability distribution function parameterized with the shape $a$ $(a>0)$ and scale $b$ $(b>0)$ parameters. Fig. \ref{fig:figure2_article2}(b) shows the geometry of 3D-RF composites with an increasing coefficient of variation $CV=1/\sqrt{a}$ used as a measure of dispersion; at $CV=20\%$, $CV=50\%$, and $CV=100\%$. The corresponding porosity is equal to $\phi=94.1\%$ and the mean fiber diameter is given by $D_m=ab=14 \mu m$. The REVs based on $D_v$ and based on $D_{iv}$ are presented in Fig. \ref{fig:figure2_article2}(c) and \ref{fig:figure2_article2}(d), respectively.
Coefficient of variation was around $40\%$ in the cotton felts case (F1-F4) and $30\%$ in the PET felts one, and the orientation of fibers of the 3D-RF composites was parameterized according to an additionally needed parameter $\Omega_{zz}$ ($0\leq\Omega_{zz}\leq1$) \cite{advani1987use}. Therefore, assuming a Gamma-based distribution of fiber diameters, geometrical parameters of the model include porosity $\phi$, mean fiber diameter $D_m$, coefficient of variation of the fiber diameters $CV$ and angular orientation of fibers $\Omega_{zz}$. There exist a strong analogy between the visco-inertial frequency-dependent response function of a Newtonian  fluid-filled rigid porous medium \cite{johnson1987theory}, and its frequency-dependent thermal counterpart \cite{champoux1991dynamic, lafarge1997dynamic}. Therefore, the $D_v$-based cell was simultaneously utilized as the REV to compute the static viscous permeability $k_0$ and the static thermal permeability $k_0'$. On the other hand, on the asymptotic high frequency range, the $D_{iv}$-based cell was used as the REV for the calculation of the viscous $\Lambda$ and thermal $\Lambda'$ characteristic lengths together with the tortuosity $\alpha_{\infty}$. The fiber diameters' dispersion effects for the $D_v$-based cell and $D_{iv}$-based cell is shown in Fig. \ref{fig:Dviv_CV}.  Figure \ref{fig:Dviv_CV} demonstrates a significant increase of the volume-weighted average diameter, $D_v$, with respect to the corresponding mean diameter $D_m$, for an  increasing coefficient of variation $CV$ of the Gamma-based law. This increase of $D_v/D_m$ can be characterized by a polynominal function of second order with respect to $CV$. Meanwhile, the inverse volume-weighted diameter $D_{iv}$ undergoes an exponential decrease as the coefficient of variation $CV$ of the fiber diameters increases, due to the increasing amount of small fibers present in the probability distribution function for high levels of $CV$ (Fig. \ref{fig:figure2_article2}a). This indicates that polydisperse 3D-RF composites having $D_v$ and $D_{iv}$ based cells as REVs can introduce drastic contrasts in the local characteristic sizes ($D_v$; $D_{iv}$) and the corresponding asymptotic transport parameters ($k_0$, $k'_0$; $\Lambda$, $\Lambda'$, $\alpha_{\infty}$) owing to their strong and different sensitivities to the extreme values of the distribution. These findings suggest that the polydisperse-based RF microstructures effectively exhibit a tuning effect for the design of sound absorbing fibrous materials.
\FloatBarrier
\subsection{Advantages}
Replacing the monodisperse fiber diameters with polydisperse 3D materials can significantly enhance the viscous $k_0$ and thermal $k'_0$ static permeabilities.  It can also reduce the viscous $\Lambda$ and thermal $\Lambda'$ characteristic lengths, increase the sound absorption average (SAA), and  creates subwavelength sound absorbers. 

\subsubsection{Tuning transport parameters}  
\begin{figure*}[ht]
    \centering
    \includegraphics[width=12cm]{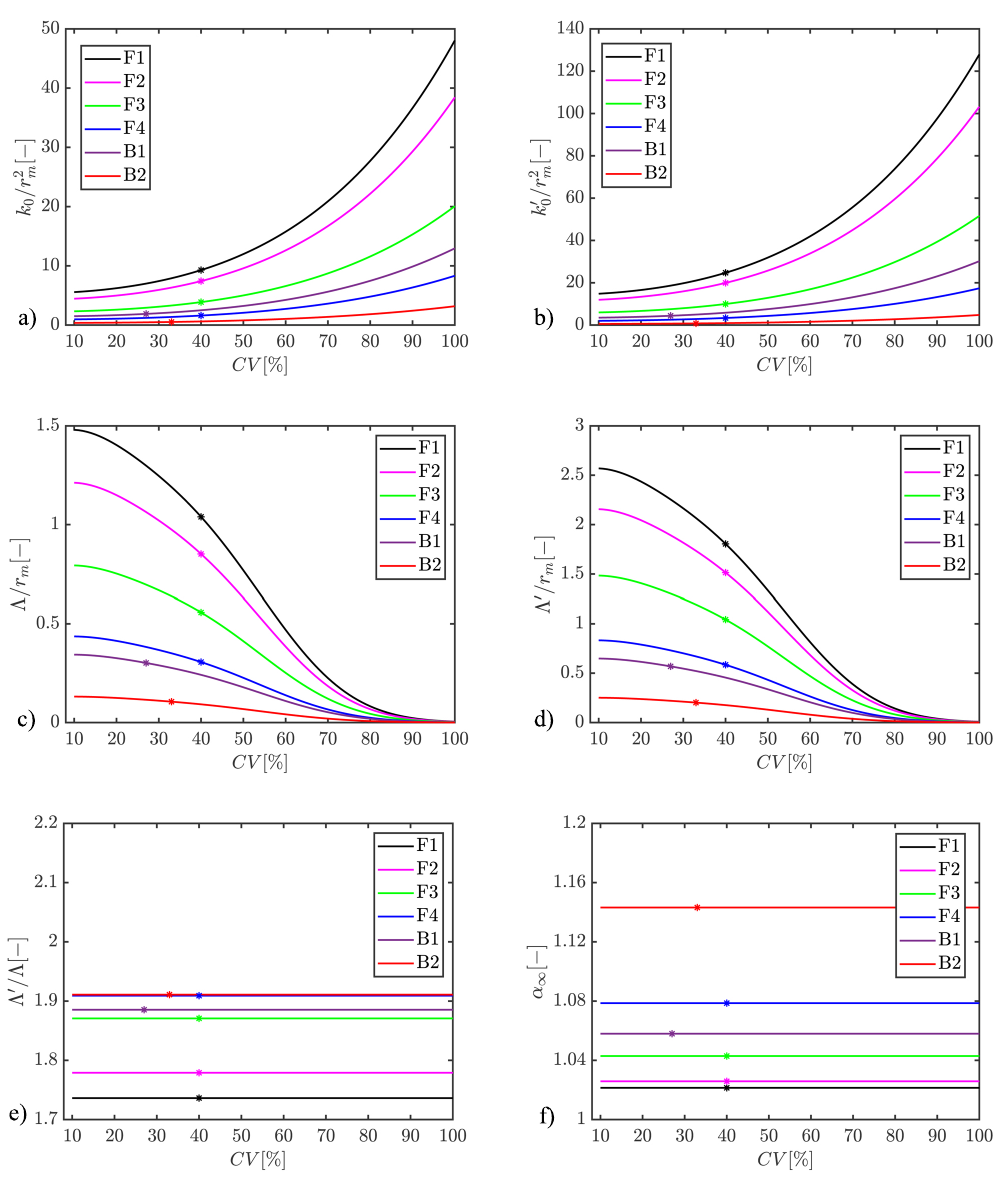}
    \caption{Dimensionless transport parameters as a function of the coefficient of variation $CV$ of fiber diameters for cotton felts (F1-F4) and PET felts (B1-B2). The porosities $\phi$ and orientation of fibers $\Omega_{zz}$ of each felt are unchanged with $CV$ and taken from measurement.\cite{tran2023effect} The dots correspond to the experimentally determined values of $CV$ in the initial state of polydispersity.}
    \label{fig:Parameter_CV}
\end{figure*}

In this study, we analyze the evolution of transport properties as a function of the coefficient of variation $CV$ for two different families of composite microstructures: composite felts (F1-F4) and PET felts (B1-B2). Figure \ref{fig:Parameter_CV} illustrates the transport parameters of these 3D-RF composites when increasing the coefficient of variation $CV$ of the fiber diameters (at unchanged porosities $\phi$ and orientation of fibers $\Omega_{zz}$). We observe a non-linear increase in the dimensionless static viscous $k_0/r^2_m$ and thermal $k'_0/r^2_m$ permeabilities (with $r_m=D_m/2$), while the dimensionless viscous $\Lambda/r_m$ and thermal $\Lambda/r_m$ characteristic lengths show a non-linear decrease with $CV$, and both the ratio $\Lambda'/\Lambda$ and the tortuosity $\alpha_{\infty}$ remain independent of the polydispersity ($\alpha_{\infty}$ is independent of the length scale of the microstructure). This suggests that controlling the polydispersity of fiber diameters through the coefficient of variation $CV$ with the manufacturing process provides a new lever for optimizing sound-absorbing fibrous materials.

\subsubsection{Maximize sound absorption average } 
\begin{figure*}[ht]
    \centering
    \includegraphics[width=12cm]{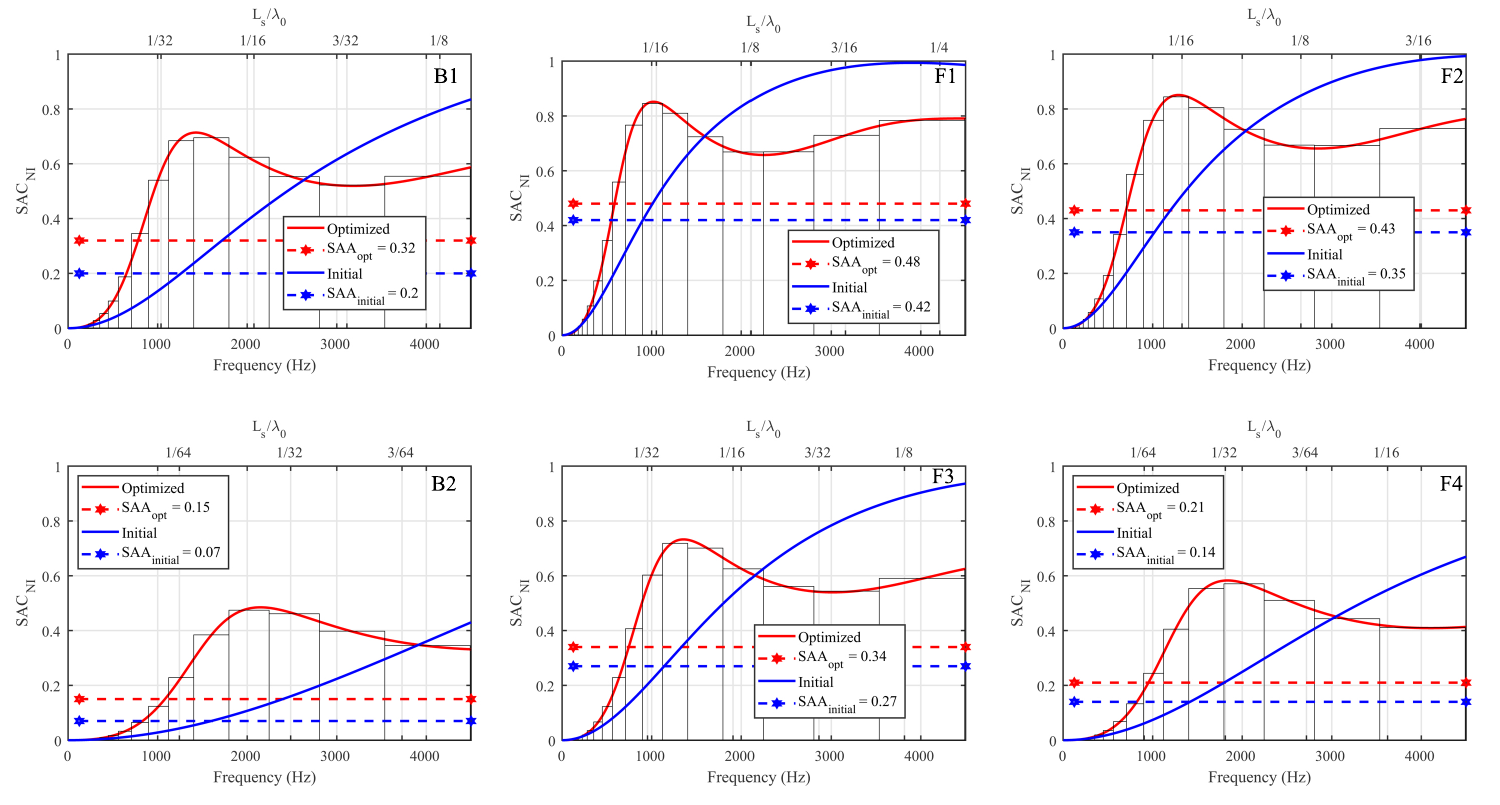}
    \caption{The sound absorbing coefficient at normal incidence, $SAA_{NI}$ as a function of frequency for the two families of felts studied in this work: Cotton felts (F1-F4) and PET felts (B1-B2). The $SAA_{NI}$ responses obtained prior to modification of the polydispersity (coefficient of variation $CV$) are shown with a blue line. The red line represents the $SAA_{NI}$ abtained after optimization of the polydispersity by maximizing the $SAA_{NI}$ between the third octave bands from 125 to 4500 $Hz$. The two dotted lines correspond to the associated single number ratings before (blue) and after (red) optimization.}
    \label{fig:SAA_125_4500}
\end{figure*}

The sound absorption average at normal incidence ($SAA_{NI}$) of a material is computed with its sound absorption coefficients at normal incidence ($SAC_{NI}$) at 16 one-third octave bands $f_i$ from 125 to 4500 $Hz$ with 
\begin{align}
    SAA^{125-4000}_{NI}=\frac{1}{16}\sum^{4500}_{f_i=125}{SAC_{NI}(f_i)}, \label{eq:SAA125_4000}
\end{align}
Here, the $SAC_{NI}$ is calculated with the material of thickness $L_s$ placed on a hard reflecting backing. Consequently, $SAA_{NI}$ is a measure of the energy dissipated by visco-thermal losses in the material for a unit incident acoustic power. This single number rating is an important coefficient that can be used for characterizing the sound absorbing properties of porous materials in a large frequency range, including low frequencies. We have previously shown that using polydispersity can significantly modify the transport properties of 3D-RF composites at constant porosity $\phi$ and angular orientation $\Omega_{zz}$. Figure \ref{fig:SAA_125_4500} shows how the sound absorption average at normal incidence $SAA^{125-4000}_{NI}$ can be maximized by optimization of the $CV$ from the initial set of studied materials (F1-F4; B1-B2). The increase in $SAA^{125-4000}_{NI}$ in 3D-RF composites is due to a larger fiber diameter polydispersity ($CV\simeq75\%$) that allows for a strong contrast between the largest fibers that dominate transport phenomena in the low frequency regime ($k_0$; $k'_0$) and the smallest fibers that govern transport phenomena in the high frequency regime ($\Lambda$, $\alpha_{\infty}$; $\Lambda'$); see Tab. \ref{tab:SAA_125_4500}. Here, the optimization was based on a differential evolution algorithm as proposed elsewhere \cite{DE_Storn1997, DE_Storn2005}.  The results show that the optimized 3D-RF materials (controlled by polydispersity) exhibit lower viscous $f_v$ and thermal $f_t$ transition frequencies than the initial materials -- which allowed visco-thermal dissipation mechanisms to occur in a lower frequency range.

In the previous calculations, the sound absorbing coefficient at normal incidence,  $SAC_{NI}$, derives from the knowledge of the intrinsic transport parameters of the 3D-RF materials ($k_0$, $k'_0$, $\Lambda$, $\Lambda'$, $\alpha_{\infty}$). The frequency-dependent response functions, $\tilde\rho_{eq}(\omega)$ and $\tilde K_{eq}(\omega)$ are described as \cite{johnson1987theory, champoux1991dynamic, lafarge1997dynamic}
\begin{align}
\begin{split}
    \label{eq:rho_w_a2}
   \tilde\rho_{eq}(\omega)=\\
  & \frac{\alpha_{\infty}\rho_{0}}{\phi}\left[1+\frac{\phi\eta}{i\omega k_0\alpha_{\infty}\rho_{0}}\sqrt{1+i\frac{4\alpha^{2}_{\infty}k_0^2\rho_{0}\omega}{\eta\Lambda^{2}\phi^{2}}}\right],
    \end{split}
\end{align}
and\\
\begin{align}
\begin{split}
    \label{eq:K_w_a2}
   \tilde K_{eq}(\omega)=\\
  & \frac{\gamma P_0/\phi}{\gamma-(\gamma-1)\left[1-i\frac{\phi\kappa}{k'_{0}C_{p}\rho_{0}\omega}\sqrt{1+i\frac{4k_{0}^{'2}C_{p}\rho_{0}\omega}{\kappa\Lambda'^{2}\phi^{2}}}\right]^{-1}},
   \end{split}
\end{align}
where $\tilde\rho_{eq}(\omega)$ is the equivalent dynamic density, $\tilde K_{eq}(\omega)$ is the equivalent bulk modulus, $\omega=2\pi f$ is the angular frequency, $i$ is the imaginary unit, $\rho_0$ is the mass density of air,  $\eta$ is the dynamic viscosity of air, $P_0$ is the atmosphere pressure,  $\gamma=C_p/C_v$ is the ratio of heat capacities at constant pressure and volume, and $\kappa$ is the heat conductivity of air.

By knowing the visco-inertial $\tilde\rho_{eq}(\omega)$ and thermal $\tilde K_{eq}(\omega)$ response functions of the equivalent fluid of the porous medium, we can determine the wave number $\tilde q_{eq}(\omega)$ and the characteristic impedance $\tilde Z_{eq}(\omega)$ of the material,
\begin{align}
    \tilde q_{eq}(\omega)=\omega \sqrt{\tilde\rho_{eq}(\omega)/\tilde K_{eq}(\omega)}, \label{eq:qw_a2}\\
    \tilde Z_{eq}(\omega)= \sqrt{\tilde\rho_{eq}(\omega)\tilde K_{eq}(\omega)}. \label{eq:Zw_a2}
\end{align}
At normal incidence, the surface impedance $ \tilde Z_s(\omega)$ and sound absorption coefficient $SAC_{NI(\omega)}$ of a porous material of thickness $L_s$, backed by a rigid and impervious wall, for any angular frequency $\omega$, are provided by
\begin{align}
   \tilde Z_s(\omega)=-i\tilde Z_{eq} \cot{(\tilde q_{eq}L_s)}, \label{eq:Zs_NI_a2}\\
     SAC_{NI}=1-\bigg |\frac{\tilde Z_s-Z_0}{\tilde Z_s+Z_0}\bigg |^2, \label{eq:SAC_NI_a2}
\end{align}
where $Z_0 = \rho_0c_0$ is the impedance of the air and $c_0$ is the sound speed in air.

In Fig. \ref{fig:SAA_125_4500}, it can be observed that there is a significant change in the wavelength-to-thickness ratio $\lambda_0/L_s$, where $\lambda_0=\omega_0/c_0$, and $\omega_0$ is the angular frequency at the first peak absorption.  In fact, optimized polydispersed-based 3D-RF composites show ratios $\lambda_0/L_s$ that can achieve values lower than $1/32$ (F4 and B2) - note that the typical ratio of conventional porous materials is around $1/4$. This result can be seen as the signature of a certain class of metamaterials called subwavelength material (or slow-sound material).  In fact, the optimized properties increase $\rho_{eq}$ and decrease $K_{eq}$.  Consequently, the sound speed in the equivalent medium is reduced. These results demonstrate that the use of controlled fiber diameter polydispersity, such as that provided with a gamma distribution shape and characterized with a $CV\sim 80\%$, has the potential to enhance the performance of sound-absorbing nonwoven materials.\\
\begin{table*}[ht]
\centering
\begin{tabular}{cccccccccccc}
\hline\hline
\multirow{2}{*}{} &     & $CV$ & $\phi$  &$k_0$ & $k'_0 $  & $\alpha_{\infty}$ & $\Lambda $ & $\Lambda'$  &$f_{v} $ &$f_{t} $  &$SAA_{NI}^{125-4500}$  \\ 
 &      & $(\%)$ & $(-)$ & $(\times 10^{-10}m^2)$ &  $(\times 10^{-10}m^2)$ & $(-)$ &  $(\mu m)$ & $(\mu m)$  & $(Hz)$ & $(Hz)$ & $(\%)$ \\ \hline
\multirow{2}{*}{F1} & Init.  &40.3  &0.948    & 4.5 & 12.1               & 1.022             & 48                & 84                    & 4668 & 2557 & 42            \\
                    & Opt. &73.3 &0.948& 11.1  & 27.8        & 1.022         & 6.9         & 12                    & 2018 & 1112 &48                    \\ \hline
\multirow{2}{*}{F2} & Init. &39.8 &0.941     & 3.9 & 9.9              & 1.026             & 42                & 74                       & 5684 & 3103 &35        \\
                    & Opt. &74.6& 0.941 & 10.0 & 27.1           & 1.026         & 5.3          & 9.4                 & 2213 & 1209 &43                     \\ \hline
\multirow{2}{*}{F3} & Init.  &41.9 &0.914     & 2.1 & 5.1              & 1.042             & 25                & 47                     & 10102 & 5850 &27 \\            
                    & Opt. &72.1 &0.914 & 4.7  & 13.1           & 1.042         & 3.3          & 6.1                &4533 & 2631   &34                    \\\hline
\multirow{2}{*}{F4} & Init.  &38.9   &0.856   & 0.7 & 1.6              & 1.077             & 15                & 28                       & 25286 & 17463  &14    \\
                    & Opt. &73.8 &0.856 & 1.9  & 5.1          & 1.077         & 1.3          & 2.4                & 9998 & 7073    &21                  \\ \hline
\multirow{2}{*}{B1} & Init. &26.6  &0.888     & 2.7 & 6.3               & 1.056             & 42                & 79                       &7215 & 4601  &20       \\
                    & Opt. &75.5&0.888 & 9.6 & 23.1           & 1.056         & 3.3          & 6.1                 &2116 & 1356   &32                \\ \hline
\multirow{2}{*}{B2} & Init. &33.1    &0.76   & 0.7 & 1.2             & 1.144             & 15                & 29                          & 21495 & 21023  &7   \\
                    & Opt. &77.6 &0.76  & 2.4  & 4.3       & 1.144         & 0.8           & 1.6                 & 6690 & 6518    &15    \\ \hline\hline
\end{tabular}
\caption{Transport parameters evolution ($\phi$, $k_0$, $k'_0$, $\Lambda$, $\Lambda'$, $\alpha_{\infty}$) before (Init.) and after (Opt.) optimization of polydispersity ($CV$) in the three-dimensional random fibrous microstructures of cotton felts (F1-F4) and PET felts (B1-B2). $f_v$ and $f_t$ represent the visco-inertial and thermal transition frequencies which have been significantly reduced after optimization, which consists in maximizing the sound absorption average at normal incidence across the 16 one-third octave bands ($f_i$) ranging from 125 to 4500 $Hz$, $SAA^{125-4000}_{NI}$ [relative increase: F1(14\%), F2(23\%), F3(26\%), F4(50\%), B1(60\%), B2(114\%)].}
\label{tab:SAA_125_4500}
\end{table*}
\subsubsection{Maximize diffuse field sound absorption} 
Sound absorbing targets present in the industry for a porous material generally corresponds to a diffuse field (DF) acoustical excitation, satisfying $SAA_{DF}\geq0.8$, for $f_i\geq f_t$, where $f_t$ is a given one-third octave band. Lowering $f_t$ can lead to various issues in designing sound absorbing materials, as absorbing low frequencies is a challenge that suggests targets of increasing difficulty: $f_t=1600 Hz$ (Target 1), $f_t=1250 Hz$ (Target 2), $f_t=1000 Hz$ (Target 3), $f_t=800 Hz$ (Target 4) and $f_t=630 Hz$ (Target 5). These targets are defined without air gap behind the materials (hard-backed configuration), and the recommended or targeted thickness and mass are usually not mentioned, suggesting that the targets need to be reached with the lowest weight and cost. The sound absorption coefficient in diffuse field, $SAC_{DF}$, is described as:
\begin{align}
    SAC_{DF}= \frac{\bigintsss_{\theta_{min}}^{\theta_{max}}SAC(\omega,\theta)\cos{\theta}\sin{\theta} d\theta}{\bigintsss_{\theta_{min}}^{\theta_{max}}\cos{\theta}\sin{\theta} d\theta},\label{eq:SAC_DF1}
\end{align}
where $\theta_{min}=0^o$ and $\theta_{max}=90^o$ are the selected diffuse field integration limits.\\
$ SAC(\omega,\theta)=1-\bigg |\frac{\tilde Z_s(\omega,\theta)\cos{\theta}-Z_0}{\tilde Z_s(\omega,\theta)\cos{\theta}+Z_0}\bigg |^2$ is the sound absorption coefficient at oblique incidence, $\tilde Z_s(\omega,\theta)=-i\tilde Z_{eq}(\tilde q_{eq}/ \tilde q_x) \cot{(\tilde q_{eq}L_s)}$ is the surface impedance at oblique incidence with $ \tilde q_x=\sqrt{\tilde q_{eq}^2 - q_t^2}$ the longitudinal wave number and $q_t=q_0\sin{\theta}$ the transverse wave number. 

\begin{figure*}[ht]
    \centering
    \includegraphics[width=12cm]{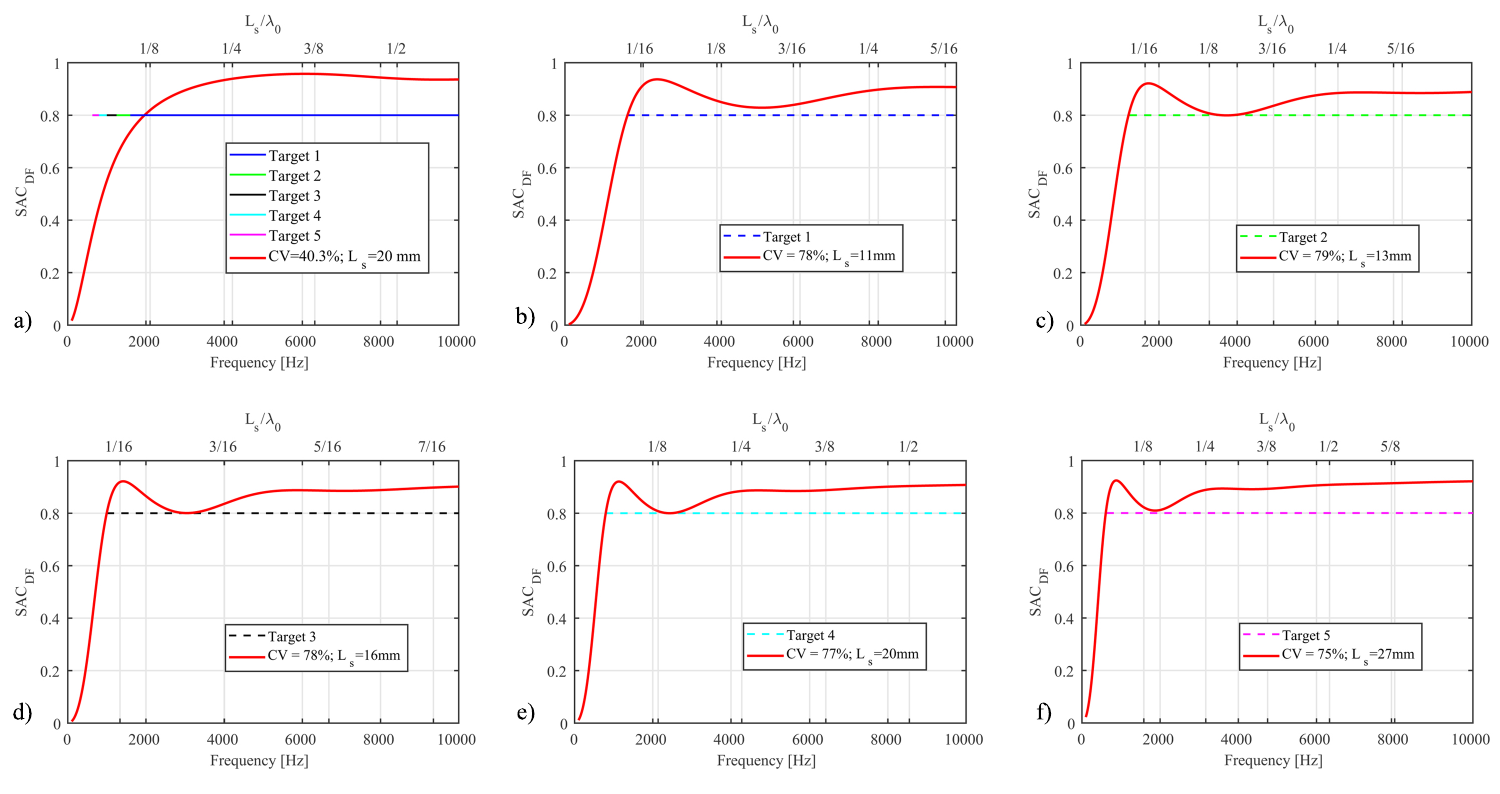}
    \caption{Diffuse field sound absorbtion coefficient $SAC_{DF}$ as a function of frequency for cotton felt F1: (a) initial configuration with $CV=40.3\%$ and $L_s=20 mm$. The next panels (b) to (f) represent the optimized configurations satisfying targets of increasing difficulty when the polydispersity degree ($CV$) is allowed to be increased as a microstructural optimization lever, for the targets listed in the legend.}
    \label{fig:Target_Industrial}
\end{figure*}

Our research demonstrates that 3D-RF composite materials, specifically polydisperse with controlled $CV\sim 75\% -79\%$ can serve as effective sound-absorbing materials in diffuse field to achieve the required industrial targets. By simulating the $SAC_{DF}$ from the initial felts (F1-F4; B1-B2), we identified the optimal coefficient of variation $CV$ allowing us to reach the targets of increasing level of difficulty with the minimal thickness. The results, shown in Fig. \ref{fig:Target_Industrial}a for the case of cotton felt F1, revealed that none of the targets can be reached from the initial microstructure ($CV=40.3\%$) and the thickness of the sample ($L_s$ = 20 mm). Fig. \ref{fig:Target_Industrial}b demonstrated an achievement of Target 1 by using a $CV$ of $78\%$ with a significant reduction of the sample thickness ($L_s$ = 11 mm). We obtained similar polydispersity results for the other targets of increasing difficulty (Target 2, $CV=79\%$, Fig. \ref{fig:Target_Industrial}c; Target 3, $CV=78\%$, Fig. \ref{fig:Target_Industrial}d; Target 4, $CV=77\%$, Fig. \ref{fig:Target_Industrial}e; Target 5, $CV=75\%$, Fig. \ref{fig:Target_Industrial}f). However, the sample thicknesses had to be increased (Target 2, $L_s$ = 13 mm, Fig. \ref{fig:Target_Industrial}c; Target 3, $L_s$ = 16 mm, Fig. \ref{fig:Target_Industrial}d; Target 4, $L_s$ = 20 mm, Fig. \ref{fig:Target_Industrial}e; Target 5 $L_s$ = 27 mm, Fig. \ref{fig:Target_Industrial}f. To better appreciate the performance that has been obtained by increasing and controlling the degree of polydispersity $P_d$ through the coefficient of variation $CV$ of the Gamma law, the results can  be plotted in terms of the wavelength-to-thickness ratio, $\lambda_0/L_s$. Using a controlled degree of polydispersity allowed us to reach sub-wavelength sound absorption in the diffuse field by lowering the maximum sound absorption peak below the quarter-wavelength resonance frequency ($L_s/\lambda_0=\frac{1}{4}$), indicating a greater ability to absorb low-frequency sound. Therefore, using 3D-RF composite materials with controlled polydispersity to efficiently absorb low-frequency sound is an important approach as it\textit{ a priori} does not require substantial modifications of the manufacturing processes. 
\FloatBarrier
\section{Outlook and perspective}
In this section, we discuss the challenges of this research area and a possible roadmap for a future work. 
\subsection{Challenges from the manufacturing process}
\begin{figure}[ht]
    \centering
    \includegraphics[width=8.5cm]{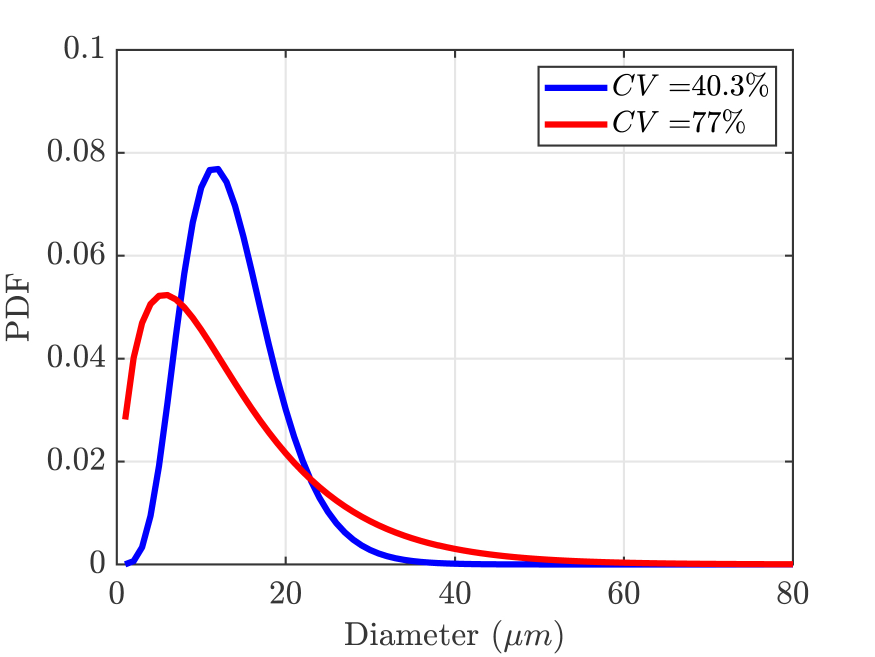}
    \caption{Comparison of the Gamma distribution of fiber diameters obtained through SEM images ($CV=40.3\%)$, cotton felt F1 and the Gamma distribution of fiber diameters required to reach Target 4 $(CV=77\%)$ }
    \label{fig:Compare_PDF_F1}
\end{figure}

3D-RF composites, including cotton felts and PET felts, are commonly manufactured by techniques coming from the textile industry. After their selection, the recycled (shoddy, PET) or noble (bicomponent) fibers are provided within different fineness specified by the $dtex$ unit which enables a linear density estimates of fiber size, such that a corresponding fiber diameter can be determined. However, the model we developed from which we evidenced a strong potential of optimization is not without drawbacks, as it relies on continuous distribution of fiber diameters. Fig. \ref{fig:Compare_PDF_F1} provides the distribution of fiber diameters obtained through SEM images that corresponds to the F1 cotton felt sample and compares this distribution with the one required to reach Target 4 ($CV=77\%$, same sample thickness). On the left side of the distribution, the fibrous material presents a relatively high number of very thin fibers. The asymmetry of the probability distribution function suggests a mix of at least three or four initial families of fibers of different linear densities ($dtex$) with lower $CV$ ($20\%-30\%$) to achieve the required distribution. However, the fiber distribution of the corresponding manufactured material may exhibit discontinuity compared to the targeted Gamma law. This discontinuity is not necessarily problematic if, as expected, the microstructural descriptors governing the physics of the transport parameters are $D_v$ and $D_{iv}$. Also, the experimental validation of the model is still incomplete, since it relies on samples already manufactured which do not correspond to the expected optimal behavior ($CV\sim 30\%$ $\neq$ $CV\sim 77\%$).
\FloatBarrier
\subsection{Future roadmap}
In order to address the challenges associated with discontinuity and validation of the model in highly polydispersed 3D-RF composites, a crucial aspect of our future roadmap involves the development of a manufacturing method that enables the precise and controlled elaboration of a relatively continuous distribution of polydisperse fiber diameters with $CV\sim 77\%$ from a chosen reduced set of fiber families with $CV\sim30\%$. 

Another promising method for achieving continuous fiber diameter distribution with a high degree of polydispersity is Melt-Spun Fibers (MSF) \cite{Rudolf2020}. The melt spinning technique is a mature method for manufacturing polymer fibers from thick filaments $\sim 100 \mu m$ to submicrometric fibers. The method is inexpensive and can offer the scalable production of 3D-RF composites as a chosen polydispersity degree $P_d$. In our forthcoming work, we aim to leverage MSF-methods on a chosen basis of polydisperse fiber diameters in our experiments to achieve improved 3D-RF composite structures with controlled distribution of fiber diameters. 
\section{Summary}
In summary, this article highlights the potential of using three-dimentional random fibrous microstructures with various distributions of fiber diameters as a way to control visco-thermal dissipation mechanisms in sound absorbing materials. This approach offers numerous advantages including tuning transport parameters across a broad range of values, enhancing the sound absorption average, achieving targets of sound absorption in diffuse field with increasing levels of difficulty, and sub-wavelength sound absorption. However, challenges persist, such as manufacturing a continuous distribution of fiber diameters with prescribed properties. To overcome this challenge, we suggest reconstructing the prescribed distribution with a limited set of three or four families of fibers carefully selected, or exploring the synthesis of a distribution of fiber diameters through the melt spinning technique. These perspectives offer promising avenues for further advancements in the field of utilizing 3D-RF composite microstructures for subwavelength sound absorbing materials.
\FloatBarrier
\section*{Acknowledgments}
This work was supported by the Association Nationale de la Recherche et de la Technologie and Adler Pelzer Group under convention Cifre No. 2020/0122 and the Natural Sciences and Engineering Research Council of Canada (Ref. RGPIN-2018-06113). We greatly acknowledge the fruitful discussions with Jean-Yves Curien during the course of this work.

\appendix

\bibliographystyle{elsarticle-num-names} 
\bibliography{References}




\end{document}